\documentclass[11pt]{article}
\usepackage{theorem}
\usepackage{latexsym}
\usepackage{amsmath}
\usepackage{amssymb}
\usepackage{amscd}
\clearpage
\usepackage[dvips]{graphicx}
\usepackage{graphicx}
\usepackage{subfigure}

\begin{document}
\baselineskip0.5cm
\renewcommand {\theequation}{\thesection.\arabic{equation}}
\title{Spectral energy distribution and generalized Wien's law for
photons, cosmic string loops and related physical objects
}

\date{}
\author{D.~Jou$^1$ \and M.S.~Mongiovi$^2$ \and M. Sciacca$^3$}
\maketitle

\begin{center}
{\footnotesize $^1$ Departament de F\'{\i}sica,  Universitat
Aut\`{o}noma de
Barcelona, Bellaterra, Catalonia, Spain\\
$^2$ DIEETCAM, Universit\`a di
Palermo, Palermo, Italy\\
$^3$ Dipartimento SAgA, Universit\`a di Palermo, Palermo,
Italy\\
E-mail addresses: david.jou@uab.es, m.stella.mongiovi@unipa.it,
michele.sciacca@unipa.it}
 PACS number(s): 98.80.Cq; 95.36.+x; 05.70.Ln; 05.70.Ce
\end{center}

\begin{abstract}

Physical objects with energy $u_w(l) \sim l^{-3w}$ with $l$ a
characteristic length and $w$ a numerical constant ($-1  \leq w  \leq
1$), lead  to an equation of state $p=w\rho$, with $p$ the
pressure and $\rho$ the energy density. Special objects with this
property are, for instance, photons ($u = hc/l$, with $l$ the
wavelength) with $w = 1/3$, and some models of cosmic string loops ($u =
(c^4/aG)l$, with $l$ the length of the loop and $a$ a numerical
constant), with $w = -1/3$, and maybe other kinds of objects as,
for instance, hypothetical cosmic membranes with lateral size $l$
and energy proportional to the area, i.e. to $l^2$, for which $w =
-2/3$, or the yet unknown constituents of dark energy, with $w =
-1$. 
Here, we discuss the general features of the spectral energy
distribution of these systems and the corresponding generalization
of Wien's law, which has the form $Tl_{mp}^{3w}=constant$,  being $l_{mp}$ the most probable size of the mentioned objects.
\end{abstract}

\section{Introduction}

The search for candidates for dark energy has been a stimulus to
recent thermodynamics of exotic systems which had not attracted
the interest of researchers before the discovery of cosmic
acceleration and the need for systems with sufficiently negative
pressure  \cite{Amendola}--\cite{Gong}. Dark energy is believed to
contain seventy per cent of the energy of the whole universe, and
to make the cosmic expansion to accelerate. The latter fact
requires that $p < -(1/3) \rho$, $\rho$ being the internal energy
density and $p$ the pressure. Thus, much interest has been focused
on systems such that $p = w \rho$, with $w$ a numerical constant,
as simplest possibilities for dark energy. For the sake of
generality we do not restrict to $w < -1/3$, but we consider the
situations with $-1 \leq w  \leq 1$ under a single physical formalism, in order to extend all the arguments to
different physical objects. In
particular, the cases with $w =1/3$ and $w = -1/3$ correspond to a
gas of photons and to a gas of linear string loops and exhibit an
interesting duality relation between themselves
\cite{DarkEnergy,Duality,Bollettino}.

Other physical objects could be small cosmic membranes of
size $l$, with energy proportional to the area, i.e. to $l^2$, corresponding to
$w = -2/3$, as we will see below. In the present paper we do not
aim to propose a particular microscopic model for dark energy, but
to deepen into the applications of thermodynamics for a family of
objects leading to an equation of state of the form $p = w \rho$,
in particular, in the application of thermodynamic ideas to obtain
information on the spectral energy distribution of such known or
hypothetical objects.

From the spectral distribution one may obtain a generalized Wien's
law relating temperature $T$ and the most probable size
characterizing the mentioned objects, which for photons reduces to
the well-known form of Wien's law for electromagnetic radiation,
but which for cosmic string loops leads to a significantly
different result. Such generalized law would provide a simple link
between the geometrical features of those objects and the
temperature of the corresponding system.

In Section 2 we summarize some previous thermodynamic results, in Section 3 we present  the adiabatic theorem and its general relation with the
spectral distribution, and in Section 4 we apply them to our
family of systems. Section 5 is devoted to conclusions and
comments; in particular, we stress relevant differences between
the systems with $w>0$ and with $w<0$.

\section{Previous thermodynamic results}
\setcounter{equation}{0}

In \cite{DarkEnergy,Duality,Bollettino}, we have proposed to
consider a family of hypothetical physical entities that are
characterized by a length $l$ and an energy $u_w(l)$, given by
\begin{equation}\label{1-1}
u_w(l) =
C_w l^{-3w}
\hskip0.3in
\hbox{with}
\hskip0.3in
C_w=h c \left( \frac{c^3}{a^2 h G}
\right)^{\frac{1-3w}{2}}
\end{equation}
as photons, for $w = 1/3$ (with $l$ the wavelength and $u =
hc/l$), cosmic strings with $w = -1/3$ ($u = (c^4/aG)l$, with $l$
the length), cosmic dust, with $w = 0$ (no characteristic length,
usually taken as dots), or cosmic membranes, with energy
proportional to their area, with $l$ the lateral size and $w =
-2/3$, or cosmic quintessence, corresponding to $w = -1$.

In (\ref{1-1}) $a$ is a numerical
constant which may depend on the model of loop, and whose value is
yet  uncertain, in the range between 1 and 10$^6$, according
to current results based on the observations of the energy peaks
of astronomical flares from gamma ray bursts and from active
galactic nuclei \cite{Abramowski-2011}--\cite{Albert-2008}.

Under two further physical hypotheses,  namely: a)
 that their absolute temperature $T$ may be related to the
average value of the internal energy as $k_BT = <u(l)>$ ($<...>$
denoting the average value over the length distribution function
of the objects) and b) that the average separation between these entities 
is proportional to their average size,  these systems lead in a direct way to $p = w
\rho$, and to a
vanishing chemical potential \cite{DarkEnergy,Duality,Bollettino}. 

Note that here we are dealing with special kind of cosmic string loops with formal 
features analogous to quantized vortices in superfluids \cite{DarkEnergy,Duality}. According to b), in an expansion at constant energy,
these loops aggregate to form longer and more separated  loops, which leads 
to a decrease of entropy. Thus, $(\partial S/\partial V)_U$ is negative, and this yields a negative 
pressure. This have not been so had we considered loops whose length was not modified in the 
expansion, but which would have kept constant length and  become more separated.
A dilute gas with these latter kind of loops would have the same equation as a system of points, instead 
of yielding a negative pressure. Thus, both features a) and b) are relevant for the definition of 
the systems considered here, which may be considered as mathematical models, independently of their actual 
physical existence. 

In \cite{DarkEnergy,Duality,Bollettino} we have focused our
attention on the explicit determination of the thermodynamic
functions of these systems, as for instance $U(T, V)$, $S(T, V)$,
$S(U, V)$, $p(U, V)$, $F(T, V)$, and so on, being $U, S, F, T$ and
$V$ internal energy, entropy, Helmholtz free energy, temperature
and volume, respectively. In particular, it was shown that
\begin{equation}\label{2-2}
\rho_w(T) = \frac{U_w(T,V)}{V} = A_w T^{(1+w)/w},
\end{equation}
with $A_w$ a constant, and the entropy 
\begin{equation}\label{2-4}
S_w(T, V) = (1 + w)B_w T^{1/w} V,        
\end{equation}
with $B_w$ another constant.

Furthermore, in \cite{Duality} we have
considered in detail the duality relations between electromagnetic
radiation ($w = 1/3$) and cosmic string loops ($w = -1/3$),
while in \cite{Bollettino} it was extended to systems with $w\neq
\pm 1/3$. In the present paper we go a step beyond  and explore in more depth the thermodynamic clues on the
spectral energy distribution, i.e $\rho_w (T, \nu)$, describing
the energy density distribution among  the several modes of
frequency $\nu$ at a given temperature $T$. This is more than an
academic exercise, as this information is essential for a deeper
physical understanding of these systems, for a comparison with
other kinds of systems, and as a step towards a statistical
physics formulation for them.

\section{Adiabatic theorem and spectral energy distribution}
\setcounter{equation}{0}

In this presentation we follow the lines  by Lima and
Alcaniz \cite{Lima_2004} in their analysis of the same problem we are
interested in, namely, to explore the spectral energy distribution
of systems with $p = w \rho$ using thermodynamic methods. However,
our approach is very different, since we have concrete (although
hypothetical) models for our systems, which will let us to have
information on the dispersion relation between frequency and
characteristic lengths. As a consequence, our conclusions about
the form of Wien's law in terms of temperature and wavelength for
different values of $w$ will be very different from those obtained
in \cite{Lima_2004}, although its expression in terms of temperature and
frequency will be the same of them.

Here, instead of  the internal energy $U_w(T,V)$ of the system
in a volume $V$ at temperature $T$, we want to find thermodynamic constraints on the spectral distribution $U_w(T, V, \nu)$
or the spectral energy density $\rho_w(T, \nu) = U_w(T, V, \nu)/V$, in such a way we may write:
\begin{equation}\label{new-1}
U_w(T, V) = \int  U_w(T, V, \nu) d \nu = \int \rho_w(T, \nu) V d \nu
\end{equation}
 with  $U_w (T, V, \nu)$ the total energy
contained in a small band of frequency $\Delta\nu$ (between $\nu- (\Delta\nu)/2$ and $\nu+ (\Delta\nu)/2$) and $\rho_w (T, \nu)$ 
the energy density per unit volume. 
Our main concern will be to find the form of $\rho_w(T, \nu)$.

We start our analysis by using the  adiabatic theorem
proved by Ehrenfest in 1917 \cite{Eherenfest},  which states that
for any reversible adiabatic change of volume inside an enclosure
at temperature $T$ at thermodynamic equilibrium, the internal energy in any frequency slot 
of width $\Delta \nu$ centered in frequency $\nu$, divided by frequency $\nu$ is constant,
 i.e.:
\begin{equation}\label{7bis}
   \frac{\rho_w(T_1, \nu_1)V_1 \Delta \nu_1}{\nu_1} = \frac{\rho_w(T_2, \nu_2)V_2 \Delta
   \nu_2}{\nu_2}.
\end{equation}
Here, it means that we consider an enclosure containing the
system defined by (\ref{1-1}) at temperature $T_1$ and focus
our attention on a band of frequencies $\Delta \nu_1$ centered on
frequency $\nu_1$. Assume that temperature $T_1$ changes to
temperature $T_2$ in an adiabatic (and reversible) volume change from $V_1$ to $V_2$.
The frequency band being considered will change from $\nu_1$ and
$\Delta\nu_1$ to $\nu_2$ and $\Delta\nu_2$, and the corresponding
energy contained in these bands will be $\rho_w(T_1, \nu_1)
\Delta\nu_1$ and $\rho_w(T_2, \nu_2) \Delta\nu_2$, respectively.

Thus, according to  (\ref{7bis}) and taking into account (\ref{2-4}) and that entropy is constant in an adiabatic and  reversible process
then one has
\begin{equation}\label{equa1}
\frac{\nu_2 \rho_w(T_1, \nu_1) \Delta \nu_1 }{\nu_1 \rho_w(T_2, \nu_2) \Delta \nu_2}=\frac{V_2}{V_1}=\frac{T_1^{1/w}}{T_2^{1/w}}
\end{equation}
On the other side, according to (\ref{2-2}), the energy density should change as 
\begin{equation}\label{equa2bis}
\frac{\sum  \rho_w(T_1, \nu_1) \Delta \nu_1 }{\sum \rho_w(T_2, \nu_2) \Delta \nu_2}=\frac{T_1^{\frac{1+w}{w}}}{T_2^{\frac{1+w}{w}}}
\end{equation}
where summation refers to the slots $\Delta \nu$ in which the frequency domain can be divided because of (\ref{equa1}).  Expression (\ref{equa2bis}) can be also written using (\ref{7bis}) and (\ref{equa1}) as
\begin{equation}\label{equa2}
\frac{\sum \left(\frac{\nu_1 T_2}{\nu_2 T_1}\right) \rho_w(T_2, \nu_2) \Delta \nu_2 }{\sum \rho_w(T_2, \nu_2) \Delta \nu_2}=1
\end{equation}
which, because of the arbitrary choice of the ensemble $\nu_1$ and $\nu_2$ and their corresponding slots $\Delta \nu_1$ and $\Delta \nu_2$, implies 
\begin{equation}\label{2-5}
\frac{\nu_1}{T_1}=\frac{\nu_2}{T_2}.
\end{equation}

From here follows also that $\Delta \nu_1/T_1=\Delta \nu_2/T_2$. Thus, one has from (\ref{equa1}) 
\begin{equation}\label{8bis}
\frac{\rho_w(T_2, \nu_2) }{\rho_w(T_1, \nu_1)}=\frac{T_2^{\frac{1}{w}}}{T_1^{\frac{1}{w}}}=\frac{ \nu_2^{1/w}}{ \nu_2^{1/w}}
\end{equation}
Multiplying and dividing the expression of $\rho_w(T, \nu)$  by $\nu^{1/w}$ an taking into account (\ref{2-5}), one may write $\rho_w(T, \nu)$ as
\begin{equation}\label{2-8}
\rho_w(T, \nu) =  \alpha_w  \nu^{1/w} \Phi_w (T/ \nu), 
\end{equation}
where $ \Phi_w(T/ \nu)$ is a function to be determined,
but which does not change in adiabatic processes because its
argument, $T/\nu$, remains constant in such change, according to
(\ref{2-5})  (all the frequency of the mode change as $T$ in the adiabatic process).   This may be considered as
the generalized Wien spectrum, because for electromagnetic
radiation it was Wien who proposed this kind of functional form.
He himself proposed an approximate expression of exponential form
for $\Phi_w(T/\nu)$, whose exact form for electromagnetic
radiation ($w=1/3$) was finally obtained by Planck in 1900, namely
\begin{equation}\label{2-9}
\rho_{electr~rad}(T,\nu) = \frac{8 \pi h}{c^3} \frac{\nu^3}{e^{ h
\nu/kT} -1},
\end{equation}
with $k_B$ Boltzmann's constant.

\begin{figure}
 \includegraphics[width=10cm]{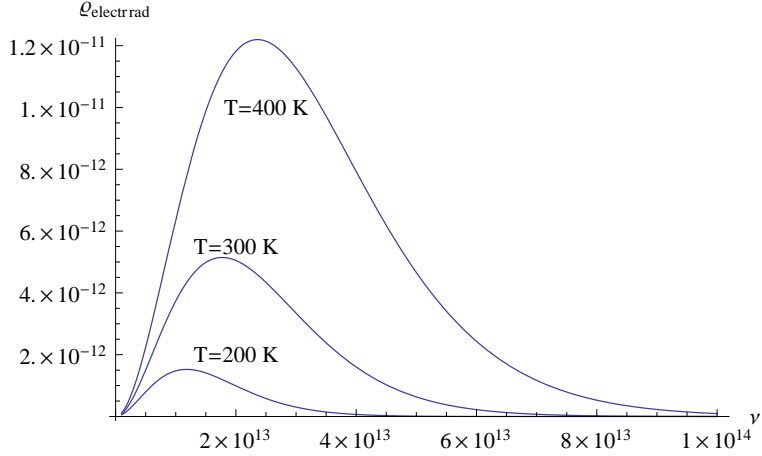}\\
\caption{\small{In this figure the expression of the energy
distribution of the electromagnetic radiation (\ref{2-9}) is
 plotted over the frequency range $\nu$. Curves are shown at three different
 temperatures ($T=200$ K, $T=300$ K and $T=400$ K)}.}\label{Planck}
\end{figure}

In the next section we turn our attention to the family of systems
defined by (\ref{1-1}), which generalizes the thermodynamics of
electromagnetic radiation to a wider family of systems.

\section{Spectral energy distribution for photons, cosmic string loops, and related physical
objects}
\setcounter{equation}{0}

To apply the ideas of the former Section to our family of systems
(\ref{1-1}) we need information on their characteristic frequency.
In the case of electromagnetic radiation, this is clear, because
$\lambda\nu  = c$, with $\lambda$ the wave-length and $c$ the speed of light in vacuum. To do
this in a more general way for all the systems in (\ref{1-1}) we
consider (\ref{1-1}) as expressions for the energy quanta in terms
of $l$ (recall that for photons (\ref{1-1}) yields $u(l)= hc/l$,
with $l$ the wavelength). Furthermore, we combine this expression
for the quanta of energy in terms of $l$ with the Einstein-Planck
expression of the energy of the quanta in terms of $\nu$, namely
$u(\nu) = h \nu$.
Then, combining (\ref{1-1}) with this expression
we obtain
\begin{equation}\label{3-1}
\nu =c \left(\frac{c^3}{ h a^2G}\right)^{\frac{1-3w}{2}} l^{-3w}.          
\end{equation}
For photons this just yields $\nu = c/l$ ($l$ being $\lambda$).
For string loops this yields
\begin{equation}\label{3-2}
\nu = \frac{c^4}{ h a^2G} l ,     
\end{equation}
which gives the characteristic frequency of a loop of length $l$.

Thus, the general form of energy spectral distribution for these
systems will be of the kind (\ref{2-8}). In principle, the form of
$\Phi_w(T, \nu)$ depends on $w$. It is tempting to assume that for
all of them an analogous of the Planck distribution holds, in such
a way that
\begin{equation}\label{3-3}
\rho_w(T, \nu) =  \frac{\alpha_w \nu^{1/w}}{e^{h\nu/kT}-1 }     
\end{equation}

But this is only a guess which turns out to be untenable for $w<0$. Indeed,
it is known \cite{DarkEnergy}--\cite{Lima_2004}  that for $w > 0$
the systems are thermodynamically stable, but that for $w < 0$
they have negative specific heat and therefore they are unstable,
and require a continuous input to keep themselves into existence
(in cosmology, this input could be the cosmic expansion itself).
Therefore, one may expect that the form of $\Phi_w(T, \nu)$ may be
considerably different for $w > 0$ and for $w < 0$. The situation
$w = 0$, corresponding to cosmic dust, implies constant energy,
independent on the length. Usually, it is taken $l = 0$ as its
characteristic length, namely, point-like dust; and the
characteristic temperature $T$ is taken as zero, in order that
$VT^{1/w}$ may be finite.

In particular, the dynamics of strings and string loops has been
studied in different systems, going from cosmic strings \cite{Vilenkin}--\cite{Davis}
to quantized vortex loops in superfluid helium \cite{Nemirovskii, JouMong-PLA-2009}. The
results indicate that one should expect a potential distribution
law. For instance, the length probability distribution in these
systems is often found to have the form \cite{JouMong-PLA-2009}
\begin{equation}\label{3-4}
n(l) dl = B_q \frac{l^{-q }}{l_{min}^{ 4-q}} dl     
\end{equation}
with $n(l) dl$ the number of loops of length comprised between $l$
and $l + dl$ per unit volume; $B_q$ is a dimensionless
normalization constant, $q$ a characteristic exponent, and
$l_{min}$ the minimum length (otherwise, for a vanishing minimum
length, the distribution function would be divergent).

But here we aim to relate (\ref{3-4}) to the spectral energy
distributions (\ref{2-8}). To do so, we need to express the energy
distribution, to relate $l$ to the frequency $\nu$, and to
introduce temperature $T$, which is not evident,
because (\ref{3-4}) differs very much from the usual forms of equilibrium
statistical distribution functions, for which the temperature is
well identified.
In Ref \cite{DarkEnergy},\cite{Duality} we have used for temperature the definition
\begin{equation}\label{3-7}
k_BT = <u(l)> ,  
\end{equation}
with $<...>$ standing for the average value of the energy over the
distribution of lengths of the objects. In Ref \cite{DarkEnergy}
we showed that using distribution (\ref{3-4}) one obtains
\begin{equation}\label{3-8}
k_B T = \frac{q-1}{q - 2 +  ({1+3w}) }  u_w(l_{min})= \frac{q-1}{q - 1 +  3w }  C_w l_{min}^{-3w}.     
\end{equation}

The energy density distribution in terms of $T$ and $l$ will be
\begin{equation}\label{3-5}
\rho_w(T, l)dl = u_w(l) n (l) dl .      
\end{equation}
We may convert it to $\rho_w(T, \nu) d\nu$ taking into account the
relation (\ref{3-1}) between $l$ and $\nu$. Since $u(\nu) = h
\nu$, we have
\begin{equation}\label{3-6}
\rho_w(T, \nu)d\nu = -\frac{1}{3w} h^{(q-1)/(3w)+1}  B_q C_w^{(1-q)/3w} \frac{ \nu^{(q-1)/3w}}{(l_{min})^{4-q}} d\nu, 
\end{equation}
where we have taken into account that $d \nu = -{3w} \frac{C_w}{h} l^{-(1 +
3w)}dl$. Equation (\ref{3-6}) for $w=-1/3$ becomes
\begin{equation}%
\rho_{-1/3}(T, \nu)d\nu =   h^{2-q}  B_q C_w^{q-1} \frac{
\nu^{1-q}}{(l_{min})^{4-q}} d\nu,%
\end{equation}
Note that this can be used only for $w <0$, otherwise the
distribution function would correspond to negative values of the
probability.

Introducing (\ref{3-8}) into (\ref{3-6}) one obtains
\begin{equation}\label{3-9}
\rho_w(T, \nu) d\nu = B'_{q,w} \nu^{ 1/w}\Phi_{ q,w}(T/\nu)d\nu          
\end{equation}
with $B'_{q,w}=-\frac{1}{3w} h^{(q-1)/(3w)+1}  B_q
C_w^{-1/w}\left(\frac{1}{k_B}\frac{q-1}{q - 1 +  3w }
\right)^{-(4-q)/3w}$
and where $\Phi_{q,w}(T/\nu)$ is
\begin{equation}\label{3-10}
\Phi_{q,w}\left(\frac{T}{\nu} \right) = \left(\frac{\nu}{T}\right)^{-(4-q)/3w} 
\end{equation}
Then, although the distribution (\ref{3-4})  does not seem to have
any relation with the energy spectral distribution (\ref{2-8}), we have seen
here that relating $l$ to energy, energy to frequency, and
introducing a suitable definition for temperature one may obtain
for the systems with $w < 0$ an energy spectral distribution
function which has indeed the form required by thermodynamic
arguments. Though this could seem natural, this is not so, because
a thermodynamic formalism for systems with $w < 0$ is scarcely
known; thus, our result provides an additional argument in favor
of the internal consistency of the thermodynamic analysis of
systems (\ref{1-1}).

\begin{figure}
  \includegraphics[width=10cm]{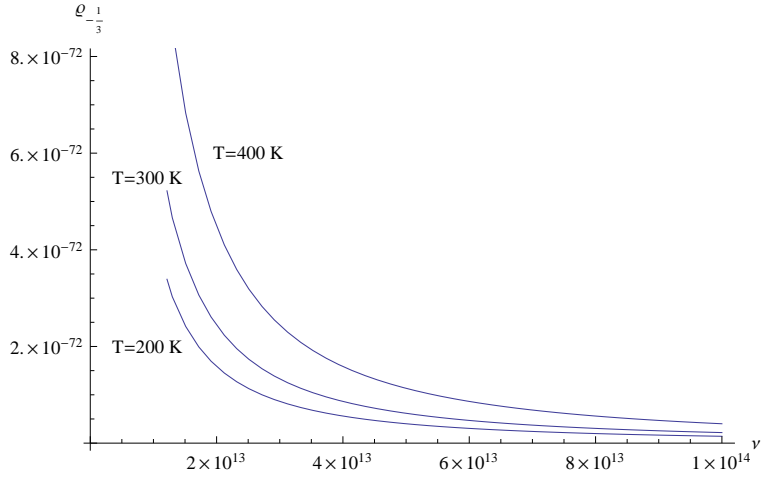}\\
  \caption{In this figure the expression (\ref{3-9}) for $w=-1/3$ is
 plotted, assuming $C_w=B_q=1$ and $q=5/2$. For a direct comparison with respect to the
 figure \ref{Planck},  curves are shown at the same three different
 temperatures($T=200$ K, $T=300$ K and $T=400$ K).
 }\label{Planck2}
\end{figure}

\section{Conclusions and remarks: generalized Wien's law for cosmic string loops
and electromagnetic radiation}
\setcounter{equation}{0}

It has been recalled that the adiabatic theorem shows that $T/\nu
= constant$ in adiabatic reversible processes. From here follows,
for instance, that the spectral distribution of $\nu$ should have
the general form (\ref{2-8}), and from here it follows, too, that
$T/\nu_{most~probable} = constant$. In the case of photons ($w =
1/3$), it follows from here the well-known Wien's law, with
$\lambda_{most~probable} T = constant$ \cite{Landau_Stat}.

Here we will comment about the generalization of Wien's law to the
case of cosmic string loops, or for cosmic membranes. Having this
generalization may be of interest to relate the temperature of
these objects to their microscopic geometrical features.

From $T/\nu_{most probable}$, one cannot conclude in general that
$\lambda T =constant$, as it has been done in Ref.\,\cite{Lima_2004}, because this is a result for photons (or for
systems with $u(l)$ proportional to $l^{-1}$). In view of
(\ref{3-1}), instead, it is seen that $T/\nu = constant$ leads to
\begin{equation}\label{4-1}
 T l_{mp}^{3w}    = constant 
\end{equation}
where $l_{mp}$ is the most probable value of the length $l$.

For electromagnetic radiation ($3w = 1$) this yields indeed
$Tl_{mp} = constant$. However, for cosmic string loops ($3w = -1$)
the corresponding form is $T/l_{mp} = constant$, for cosmic
membranes ($3w = -2$) (\ref{4-1}) yields $T/l^2 = constant$, and
for dark energy ($w = -1$) one has $T/l_{mp}^3 = constant$. Thus,
having some detailed information on the dispersion relation of the
systems with $w$ different from $-1/3$ is essential to go from the
general form $T/\nu = constant$ to the several particular forms
summarized in $Tl_{mp}^{ 3w} = constant$ in terms of the
characteristic length. Since in Ref \cite{Lima_2004} this
information was lacking, it was assumed that the form $Tl_{mp}  =
constant$ was valid for systems with all values of $w$.
In fact the behavior  (\ref{4-1}) is consistent with expansion 
(\ref{2-4}) for the entropy. In an expansion, electromagnetic radiation, cools as $(RT)^3=constant$ 
($R$ being the length scale factor of the container),
whereas the system of cosmic string loops becomes hotter, as $R/T=constant$.

Second, it is worth to remark that systems with $w < 0$ do not
seem to follow Planck statistics (nor Bose-Einstein nor
Fermi-Dirac ones). This may be related to the fact that they are
not stable equilibrium systems, but steady states kept by an
external input, which makes them to depend strongly on the
particular dynamics, which is reflected in (\ref{3-10}) through
the value of the exponent $q$. This makes even more surprising the
fact that even for these systems the energy spectral distribution
may be written in the general form (\ref{2-8}).

One topic of discussion in connection with systems with negative
pressure is about the possibility or impossibility of the
so-called phantom dark-energy. In the present family
of models, phantom dark energy with $w < -1$ would lead to a
negative entropy. Equation (\ref{3-8}) gives some more requirements
related to the distribution function (\ref{3-4}), according to which,
to have $T > 0$, one should have that the absolute value of $3w$
should be less than $q -1$, $q$ being the exponent in (\ref{3-4}). The
situation with $w = -1$ would then require that $q > 4$, but not
much is known about the actual value of $q$. Let us mention that
in superfluid turbulence, the exponent $q$ describing the length
probability distribution of quantized vortices is $q = 5/2$; in
the hypothetical case that $q$ had also this value (this is only
meant for the sake of a concrete illustration), the values of $w$
consistent with $T > 0$ would be $w > -1/2$. Thus, for the family
of systems studied here, the admissible values of $w$ are related
to their spectral properties (as for instance (\ref{3-4}) or (\ref{3-10})) in
a restrictive way.

\subsection*{Acknowledgements} The authors acknowledge the support of the Universit\`{a}
di Palermo (Fondi 60\% 2012-ATE-0106 and Progetto CoRI 2012, Azione d) and the collaboration agreement
between Universit\`{a} di Palermo and Universit\`{a}t Aut\`{o}noma
de Barcelona.
DJ acknowledges the financial support from the Direcci\'{o}n
General de Investigaci\'{o}n of the Spanish Ministry of Education
under grant FIS2009-13370-C02-01
 and of the Direcci\'{o} General de Recerca of
the Generalitat of Catalonia, under grant 2009 SGR-00164.
M.S. acknowledges the hospitality of the "Group of Fisica Estadistica of the Universit\`{a}t Aut\`{o}noma de Barcelona".

\par

\end{document}